\documentstyle{BSAXwork}

\input epsf.sty

\def \AAP #1 #2 {{\em Astron. Astrophys.\/} {\bf #1}, #2}
\def \AAL #1 #2 {{\em Astron. Astrophys. Lett.\/} {\bf #1}, L#2}
\def \AAR #1 #2 {{\em Astron. Astrophys. Rev.\/} {\bf #1}, #2}
\def \AAS #1 #2 {{\em Astron. Astrophys. Suppl. Ser.\/} {\bf #1}, #2}
\def \AJ #1 #2 {{\em Astron. J.\/} {\bf #1}, #2}
\def \ANNREV #1 #2 {{\em Ann. Rev. Astron. Astrophys.\/} {\bf #1}, #2}
\def \APJ #1 #2 {{\em Astrophys. J.\/} {\bf #1}, #2}
\def \APJL #1 #2 {{\em Astrophys. J. Lett.\/} {\bf #1}, L#2}
\def \APJS #1 #2 {{\em Astrophys. J. Suppl.\/} {\bf #1}, #2}
\def \APSS #1 #2 {{\em Astrophys. Space Sci.\/} {\bf #1}, #2}
\def \ASR #1 #2 {{\em Adv. Space Res.\/} {\bf #1}, #2}
\def \BAIC #1 #2 {{\em Bull. Astron. Inst. Czechosl.\/} {\bf #1}, #2}
\def \JSQRT #1 #2 {{\em J. Quant. Spectrosc. Radiat. Transfer\/} {\bf #1}, #2}
\def \MN #1 #2 {{\em Mon. Not. R. Astr. Soc.\/} {\bf #1}, #2}
\def \MEM #1 #2 {{\em Mem. R. Astr. Soc.\/} {\bf #1}, #2}
\def \PLR #1 #2 {{\em Phys. Lett. Rev.\/} {\bf #1}, #2}
\def \PASJ #1 #2 {{\em Publ. Astron. Soc. Japan\/} {\bf #1}, #2}
\def \PASP #1 #2 {{\em Publ. Astr. Soc. Pacific\/} {\bf #1}, #2}
\def \NAT #1 #2 {{\em Nature\/} {\bf #1}, #2}
\def \SAIT #1 #2 {{\em Mem.\ Soc.\ Astron.\ It.\/} {\bf #1}, #2}
\def \MESS #1 #2 {{\em The Messenger\/} {\bf #1}, #2}
\def \ASTRNACH #1 #2 {{\em Astron. Nach.\/} {\bf #1}, #2}

\begin{opening}

\title{Blazars, Gamma Ray Bursts and Galactic superluminal sources}
\author{Gabriele Ghisellini$^1$ and Annalisa Celotti$^2$}
\institute{$^1$Osservatorio Astronomico di Brera, Via Bianchi 46, I--23807 Merate, Italy\\
$^2$SISSA/ISAS, Via Beirut 2--4, I--34014  Trieste, Italy}
\date{} 
\end{opening}

\begin{document}

\oddpagefooter{}{}{} 
\evenpagefooter{}{}{} 
\medskip  

\begin{abstract} 
{\it Beppo}SAX is best known for its role in the disclosure of the
Gamma--Ray Burst mystery, but it has also improved our understanding
of jetted sources in general, and blazars in particular.  
On the interpretational side, we are curious to see if all sources
with relativistic jets (or ``flying pancakes", as in GRBs) are
controlled by the same basic physics, despite the very different
lifetimes and behavior.  
To this end we explore some general characteristics of blazars, GRBs
and superluminal galactic sources, such as their bulk Lorentz factors,
the power of their jets compared with what they can extract through
accretion, and the value of the magnetic field, close to the black
hole, needed to extract the spin energy of a rotating black hole.  
We find remarkable similarities, namely that the outflowing mass rate
is of the order of 1 per cent of the accretion mass rate in all
systems, and that the value of the magnetic field required to
efficiently extract the spin energy of the black hole is of the order
of the gravitational energy density of the matter close to the
gravitational radius.  
We then go on to discuss the way in which the energy in bulk
relativistic motion can be transformed into beamed radiation, and 
consider the possibility that all three classes of sources
could work in the same way, namely by a intermittent release of
relativistic plasma at the base of the jet and thus with 
similar efficiency.  
Different patches of material, with slightly different velocities,
collide at some distance from the black hole, producing the radiation
we see.

\end{abstract}

\medskip

\section{Introduction}

Radio--loud active galactic nuclei, galactic sources showing
apparent superluminal motions (GS), and gamma--ray bursts (GRBs) are
the three classes of sources where there is evidence for
large quantity of matter moving very close to the speed of light.
It is natural to wonder if these sources work in a similar way,
namely if the machine able to produce, collimate
and accelerate matter to relativistic velocities is the same.
Furthermore one would like to know if in all three systems the 
radiation we see comes from the transformation of bulk kinetic 
(or magnetic) into random 
energy and, if so, if the mechanisms able to do that are similar.
If this turns out to be true, then the galactic superluminal
sources could be used as a sort of laboratory for the study of 
radio--loud AGNs, both because their vicinity (i.e. high fluxes)
allows more detailed
studies, but also because the ratio in black hole masses is
probably $10^8$: if all timescales scale with the black hole mass, a 
millisecond in the life of GRS 1915+105 is equivalent to one day in 
the life of a blazars (see the recent review by Blandford 2002).

GRBs, on the other hand, are the most powerful events occurring after 
the Big Bang.
But this {\it does not} mean that we are witnessing a single
tremendous explosion.
Indeed, GRBs have durations lasting up to hundreds of seconds,
and recent modeling associates this time with the
duration of the central engine powering them.
Since a stellar mass black hole is involved, we have that 
a GRB of 100 s duration lasts for $\sim$a million of dynamical times.
Put in this way, the ``explosion" is instead a quasi steady--state
process!

Thus the comparison among the three systems and their central engines
is certainly legitimate, hopefully leading to 
a deeper understanding of how the formation, launch and
acceleration of jet work.



\section{The bulk Lorentz factor}

{\bf Blazars ---} The best evidences for bulk relativistic motion
in blazars come from VLBI observations of
knots of radio emission moving superluminally.
Apparent speeds up to $\beta_{\rm app} =30 h^{-1}$  
are measured (Jorstad et al. 2001, see also these proceedings),
at least in those blazars that have been discovered by EGRET to
be powerful $\gamma$--ray emitters.
With $H_0=65$ Mpc km$^{-1}$ s$^{-1}$, we have objects
moving with $\beta_{\rm app} \sim 40$, which indicates
even larger Lorentz factors $\Gamma$ (note that $\Gamma$ must always
be larger than $\beta_{\rm app}$).
These values are larger than the ``fiducial" $\Gamma\sim 10$
we are used to, and can help solving the problem of
the large radio brightness temperature of intraday variable sources
(see Wagner \& Witzel 1995 for a review),
once corrected for interstellar scintillation (sources scintillate
if their angular dimensions are small, hence the limit on the
brightness temperature; Kedziora--Chudczer et al. 2001).
Indications of a large degree of beaming come also from
spectral fitting, especially of low powerful, TeV emitting, blazars
where beaming factors $\delta>20$ are derived (Tavecchio et al. 1998). 
There are also indications that the jet could be structured, 
with a fast ``spine" surrounded by a slower ``layer"
(Laing 1993, Chiaberge et al. 2001), explaining for example
why the non--thermal radiation
of the core of radio--galaxies is not as faint as predicted if
the plasma is moving at the same large speed and if they are 
observed at large angles with respect to the jet axis.

\vskip 0.3 true cm
\noindent
{\bf Galactic superluminals ---} 
Besides SS 433 ($\beta=0.26$), GRS 1915+105 
and GRO J1655--40 (both of them have $\beta\sim 0.9$ and $\Gamma\sim$2.5)
there are other recently discovered sources showing radio jets 
with moving features (see the reviews by Mirabel \& Rodriguez 1999
and by Fender 2001). 
Among these, SAX J1819.3--2525 seems particularly interesting, since a new
determination of its distance put this source at more than 8 kpc,
much further away than previously thought (Orosz et al. 2001).
If one assumes that the jets were ejected during one of the
bright X--ray flares one then finds $\Gamma\sim 10$.

\vskip 0.3 true cm
\noindent
{\bf Gamma--Ray Bursts ---}
We do not have direct measurement of $\Gamma$ during the prompt emission, 
so we must rely on theory to estimate it:
the ``fiducial" value is $\Gamma\sim$100--300.
It cannot be much less than 100 to explain the very fast (millisecond)
variability observed in the $\gamma$--ray light curve of their
prompt emission.
It cannot be much larger than that since otherwise we would have 
that the source (the fireball) becomes transparent to Thomson
scattering before the acceleration phase ends, letting the internal
radiation escape with a quasi black body spectrum (which is not observed,
see Piran 1999 for a review).
A similar constraint ($\Gamma>100$) comes also from associating the
emission above 100 MeV seen in  a few cases with the prompt emission
(see Fishman \& Meegan 1995 for a review), thus imposing that 
the source must be transparent for the $\gamma \gamma \to e^+e^-$ process.
Similarly to blazars, there is the possibility that also the ``jets"
of GRBs are structured, with the bulk Lorentz factor
decreasing for increasing angles from the jet axis
(Rossi, Lazzati \& Rees, 2002).

\section{The power of jets}

{\bf Blazars ---} Without any doubt, the knowledge of the power and
the energy carried by jets is the prime parameter to start any
modeling.  
Despite this, the power of blazar (and radiogalaxy) jets is poorly known.  
This is due to two main reasons: 1) the radiation produced by relativistic
jets is highly beamed, and the flux we receive is
therefore highly enhanced (in blazars) or dimmed (in radiogalaxies) by
relativistic effects; 2) we still do not know for sure the matter
content of jets, i.e. we are still debating whether if
most of the matter, is made by electron--positron pairs 
or by a normal electron--proton plasma.  
On the other hand, we know that the radio lobes of radio--galaxies and
blazars are a sort of calorimeter, since the cooling time are long on
these scales, and minimum energy considerations allow to estimate 
lower limits on the total energetics.  
Yet again, even in this case we do not know either the contribution of
protons to the total energy or the emitting plasma filling factor (see 
however recent results due to Chandra observations, Fabian et al. 2002).  
Bearing in mind these limitations, the total energy of a radio lobe,
divided by its lifetime (estimated from spectral aging or from advance
motion) allows to estimate the average power required by the radio
lobes to emit and expand. 
This is the average power a jet must have.  This estimate has been
done, among others, by Rawling \& Saunders (1991): they find an
average power ranging from $10^{43}$--$10^{44}$ erg s$^{-1}$ for FR I
radiogalaxies to $10^{46}$--$10^{47}$ erg s$^{-1}$ for FR II
radiogalaxies and radio--loud quasars.

One can also calculate the power carried by the jet by
inferring its density through modeling the observed SED
and requiring that the jet carries at least the particles
and the magnetic field necessary to make the radiation we see.
This has been done on the pc scale by Celotti \& Fabian (1993),
on sub--pc scale (the $\gamma$--ray emitting zone) by Celotti \& Ghisellini
(2002, see also Ghisellini \& Celotti 2002), 
and on the hundreds of kpc scale (the X--ray jets
seen by Chandra) by Celotti, Ghisellini \& Chiaberge (2001)
and Tavecchio et al. (2000).
These studies suggest large values of the power transported
by the jet and require
the presence of a dynamically dominating proton component
(see also arguments by Sikora \& Madejski 2000).

\vskip 0.3 true cm
\noindent
{\bf Galactic superluminals ---} 
The jet of Galactic superluminal sources 
is not a steady feature, and the conditions for the
sporadic launching of jets in GRS 1915+105 and GRO J1655--40
are still under debate.
But during the major flare events in GRS 1915+105
we can calculate the jet power by knowing the bulk Lorentz
factor, the viewing angle, and assuming that the jet
is at least carrying the particles and magnetic field that 
account for the synchrotron emission we see. 
Therefore, besides the bulk motion of the emitting
particles, one has to account also for some of the 
jet power being in the form of a Poynting vector.
We therefore have a powerful tool to find the {\it minimum}
jet power, which corresponds to rough equipartition between 
the magnetic field and the particle density in the comoving frame.
In this way Gliozzi, Bodo \& Ghisellini (1999) estimated
a {\it minimum} jet power in GRS 1915+105 ranging between
$10^{39}$ erg s$^{-1}$ (pure electron--positron jet) and 
$10^{40}$ erg s$^{-1}$ (proton--electron jet).

\vskip 0.3 true cm
\noindent
{\bf Gamma--Ray Bursts ---}
Are GRB collimated?
If their emission were isotropic, then some of them
would emit more than $10^{54}$ erg only in $\gamma$--rays.
This is equivalent to one solar mass completely transformed
into energy.
One should not be particularly impressed by that, since the
rotational extractable energy of black hole is 29 per cent of its mass,
and therefore reaches $5\times 10^{54}$ erg for a 10 $M_\odot$  black hole.
On the other hand we do have some indications from afterglows 
that most of the emission
is collimated in a cone of semiaperture $\theta_{\rm j}$, 
and that therefore the true radiated
energy is a factor $(1-\cos\theta_{\rm j})$ 
less than the isotropic equivalent (for two jets).
The precise amount of collimation is uncertain (Frail et al. 2001, 
Ghisellini et al. 2002), and we do not know yet the value for 
the efficiency in transforming bulk energy into radiation.
Most energy is however contained in the first prompt phase,
and not in the much longer afterglow.
Note that in the case of GRBs the word ``jet" is somewhat
misleading, since the radial extension of the ``jet" is
of order of hundreds of light seconds.
This is the reason why Piran proposed the term "flying pancakes".
Bearing the above uncertainties in mind, the energetics of
GRBs is in the range $10^{51}$--$10^{53}$ ergs.

\subsection{Comparison with accretion disk power}

{\bf Blazars ---}
The luminosity produced by the accretion disk in blazars
$L_{\rm d}$ can be derived ``directly" only when the blue bump is visible,
i.e. when it is not completely swamped by the beamed non--thermal 
continuum or by the optical emission of the host galaxy.
In many (but not all) cases we can however estimate it
using the emission lines, both narrow (e.g. Rawlings \& Saunders 1991) and
broad (Celotti, Padovani \& Ghisellini 1997).
In completely featureless BL Lacs, instead, we can derive only
upper limits for the luminosity of the disk.
The most remarkable fact of these studies is that there is a rough
equality between the average power of the jet and the power 
in the accretion disk, once an average covering factor of
$\sim 10^{-2}$ and $10^{-1}$ is assumed for the narrow and
broad line clouds, respectively.

\vskip 0.3 true cm
\noindent
{\bf Galactic superluminals ---} 
Here the disk luminosity is emitted at relatively soft X--ray
energies, where photoelectric absorption can be an issue
(possibly making SS 433 not so bright in X--rays).
For GRS 1915+105 the X--ray luminosity, during flares,
can be $10^{39}$ erg s$^{-1}$ (see e.g. Mirabel \& Rodriguez 1999).
The recent determination of the mass of its black hole
(i.e. 14$\pm4$ solar masses, Greiner 2001) makes the 
accretion disk of GRS 1915+105 Eddington limited.
The ratio $L_{\rm j}/L_{\rm d}$ of the jet power to accretion
luminosity is then of the order of 1--10 during major ejection events.

\vskip 0.3 true cm
\noindent
{\bf Gamma--Ray Bursts ---}
In GRBs the power due to accretion is completely unobservable.
We must rely on theory and models to infer it.
There is a growing consensus that long GBRs (i.e. GRBs
whose prompt emission lasts for more than 2 seconds: these
are the only ones for which the good locations allowed
the follow--up observations in other bands)
are associated with the collapse of stars more massive than
ordinary supernovae (and therefore called hypernovae),
or a two--step collapse forming first a netron star 
and then a black hole (the SupraNova scenario of Vietri \& Stella 1998).
In any case the final resulting compact object is 
a fast spinning black hole surrounded by a very dense torus
of mass $M_{\rm T}\sim$0.1--0.2$M_\odot$.
It is possible that the duration of the GRB prompt emission is
associated with the accretion time.
If this is the case, the accretion rate is obviously huge 
even if the radiation that can escape is a tiny fraction,
since the accreting material is completely opaque
(scattering optical depths of order $10^{12}$ or so, see Table 1).


\section{Outflowing vs accreting mass rates}
For powerful blazars and Galactic Superluminals we can write:
\begin{equation}
L_{\rm d} \, =\, \eta \dot M_{\rm in} c^2
\end{equation}
\begin{equation}
L_{\rm j} \, =\, \Gamma \dot M_{\rm out} c^2
\end{equation}
and therefore
\begin{equation}
{\rm Blazars, GS}\qquad
{\dot M_{\rm out}\over \dot  M_{\rm in}}\, =\,
{\eta\over \Gamma}\, { L_{\rm j} \over L_{\rm d}}\, =\, 
10^{-2}\, {\eta_{-1}\over \Gamma_1}\, { L_{\rm j} \over L_{\rm d}}
\end{equation}
where $\eta$ is the efficiency in converting the accreted mass into
energy, $\dot M_{\rm in}$ and $\dot M_{\rm out}$
are the accretion and the mass outflowing rates, respectively. 
We use the notation $Q=10^xQ_x$.

As mentioned, in GRBs we cannot estimate the mass accretion rate directly.
Let us assume that the accretion process lasts for the duration
of the burst, and that the total accreted matter corresponds 
to the mass of the torus surrounding the black hole,
$M_{\rm T} = 0.1 M_{\rm T,-1} M_\odot$.
In this case
\begin{equation}
\dot M_{\rm in}\, =\, 2\times 10^{32} {M_{\rm T,-1}\over t_{\rm burst}}\, \, 
{\rm g~s^{-1}}
\end{equation}
yielding
\begin{equation}
{\rm GRB}\qquad
{\dot M_{\rm out}\over \dot  M_{\rm in}}\, =\,
{ L_{\rm j} t_{\rm burst}\over \Gamma M_{\rm T} c^2} \, =\,
5\times 10^{-3} {E_{\rm burst, 52} \over \Gamma_2 M_{\rm T, -1}}.
\end{equation}
We then obtain the remarkable result that (at least in this model)
the ratio of outflowing and accreting mass is roughly the same
in blazars, Galactic Superluminals and Gamma Ray bursts.
{\it If true, this result could explain why GRBs fireballs
have that particular barion loading}, which allows them to be
at the same time light enough to be relativistic
(contrary to supernovae), but heavy enough to convert all
internal energy into bulk motion (requiring the fireball
not to become transparent too early for scattering).
Alternatively, one can {\it assume} that there is a typical
(and fixed) barion loading in jets, thus deriving the resulting
bulk Lorentz factor.


\section{The central engine}

There is some consensus (although certainly not unanimous)
that the powering mechanism of jets
is the extraction of rotational energy of a spinning black hole,
through the Blandford \& Znajek (1977) process.
As an order of magnitude estimate, the power which can
be extracted by this mechanism is (Blandford \& Znajek 1977;
Rees et al. 1982)
\begin{eqnarray}
{\rm Blazars}\qquad &L_{BZ}&\, \sim \, 6\times 10^{46} 
\left({ a\over m}\right)^2 M_9^2 B_4^2 \,\,\,\,
{\rm erg~s^{-1}}\nonumber\\ 
{\rm GRBs}\qquad &L_{BZ}&\, \sim \, 6\times 10^{50} 
\left({ a\over m}\right)^2 M_1^2 B_{14}^2 \,\,\,\,
{\rm erg~s^{-1}}\nonumber\\ 
{\rm GS} \qquad &L_{BZ}&\, \sim \, 6\times 10^{38} 
\left({ a\over m}\right)^2 M_1^2 B_8^2 \,\,\,\,
{\rm erg~s^{-1}}
\end{eqnarray}
where $(a/m)$ is the specific black 
hole angular momentum ($\sim 1$ for maximally rotating black holes).

In the following we will compare the $B$--values required 
for the Blandford \& Znajek mechanism to work
with the gravitational energy density of the matter at $R_s$.
To do that, we will assume that the accretion process, in blazars and
in Galactic Superluminals, is converting in radiation a fraction 
$\eta=0.1\eta_{-1}$ of the total accreted mass--energy:
$L_{\rm d}=\eta\dot M_{\rm in} c^2$.
This allows to estimate $\dot M_{\rm in}$ from the blue bump
luminosity, or, when this is not visible,
through the luminosity of the emission lines, (assuming they reprocess
a fraction of $L_{\rm d}$ of the same order of the covering factor of
the emission line clouds, i.e. 10\% for the broad and 1\% for the narrow
emission lines, respectively).

For GRBs, instead, the mass accretion rate can be derived assuming,
as a working  hypothesis, that their duration is controlled by the
time needed to accrete the mass of the dense torus surrounding the 
spinning black hole.
In this way $\dot M_{\rm in}\sim M_{\rm T}/t_{\rm burst}$.

To find the corresponding density $n_{\rm p}$ of the matter we use
\begin{equation} 
\dot M_{\rm in}\, =\, 2\pi R h \beta_{\rm r} n_{\rm p} m_{\rm p}  c 
\end{equation}
where $\beta_{\rm r}c$ is the radial inflow velocity, and $h$ 
is the height of the disk at the radius $R$.
The gravitational energy density ${\cal E}$, 
close to the Schwartzschild radius is
\begin{equation}
{\cal E} \, \sim \, n_{\rm p} m_{\rm p} c^2\, =\, 
{\dot M_{\rm in}  c\over 2\pi (R h/R_{\rm s}^2)R_{\rm s}^2 \beta_{\rm r}  }
\end{equation}
Equipartition of the magnetic energy density 
[$B_{\rm eq}=(8\pi {\cal E})^{1/2}$] gives
\begin{eqnarray}
{\rm Blazars, GS}\qquad B_{\rm eq}\, &=& \, \left[ { 4 L_{\rm d}  \over 
        \eta c (Rh/R_{\rm s}^2)R_{\rm s}^2\beta_{\rm r} } \right]^{1/2} 
         \nonumber\\ 
{\rm GRBs}\qquad               B_{\rm eq}\, &=& \, \left[  { 4M_{\rm T} c \over 
      t_{\rm burst} (Rh/R_{\rm s}^2)R_{\rm s}^2\beta_{\rm r}}\right]^{1/2}
\end{eqnarray}
the fiducial values of $B_{\rm eq}$ are listed in Table 1 for the
three classes of sources.
One can see that the derived values are of the order required by the 
Blandford--Znajek process to provide the right power.

This can be taken as promising, since the found values match 
the required ones: note that it requires that the magnetic field
energy density is of order of the gravitational energy density,
which, depending on the accretion conditions, can be much larger than
either the gas pressure of the accreting matter or the radiation energy
density produced by accretion.
This latter possibility appears to occur for GRBs.



In Table 1 we also list the value of the scattering optical depth,
calculated in the vertical direction:
\begin{equation}
\tau_{\rm T}\, \equiv \, \sigma_{\rm T} n_{\rm p} h.
\end{equation}
We can see that in blazars it is of order unity, 
a few in (flaring) Galactic superluminals and huge in GRBs.

\begin{table}[h]
\caption{Fiducial Quantities. We use the notation $Q=10^x Q_x$.}
\begin{center}
\begin{tabular}{|l|l|l|l|l|}
\hline
 & & & & \\
Parameter &Blazars &Gal. Sup. &GRBs &Units  \\
 & & & & \\
\hline
 & & & & \\
$\Gamma$            &10--30         &2.5--10?             &$>100$                &\\
 & & & & \\
Mass                &$10^8$--$10^9$ &$\sim$10             &$\sim10$              &$M_\odot$  \\
 & & & & \\
$E_{\rm rot, max}$  &$5\times 10^{62}M_9$ &$5\times 10^{54}M_1$ &$5\times 10^{54}M_1$ &erg\\
 & & & & \\
$L_{\rm j}$       &$10^{43}$--$10^{48}$ &$10^{38}$--$10^{40}$(flare)   &$10^{50}$--$10^{52}$(iso) &erg s$^{-1}$\\
 & & & & \\
$L_{\rm d}$      &$10^{42}$--$10^{47}$ &$10^{38}$--$10^{39}$           & ?           &erg s$^{-1}$\\
 & & & & \\
$t_{\rm life}=E/L_{\rm j}$ &$5\times 10^{15}M_9 L^{-1}_{\rm j,47}$ &$5\times 10^{15}M_1 L^{-1}_{\rm j, 39}$ 
  &$500 \, M_1 L^{-1}_{\rm j, 52}$ &s  \\
 & & & & \\
$t_{\rm life}/(R_{\rm s}/c)$  &$5\times 10^{11} L^{-1}_{\rm j, 47}$ &$5\times 10^{19} L^{-1}_{\rm j, 39}$ 
  &$5\times 10^6  L^{-1}_{\rm j, 52}$ &  \\
 & & & & \\
$n_{\rm p}$ (@$R_{\rm s}$) & $4\times 10^9 {L_{\rm d, 46} (R_{\rm s}/h) \over \eta_{-1} M_9 \beta_{\rm r} }$ 
& $4\times 10^{18} {L_{\rm d, 39} (R_{\rm s}/h)\over \eta_{-1} M_1 \beta_{\rm r} }$ 
& $7\times 10^{29} {M_{\rm T,-1} (R_{\rm s}/h)\over t_2 M_1^2 \beta_{\rm r} }$ &cm$^{-3}$ \\
 & & & & \\
$\tau_{_{\rm T}}$(@$R_{\rm s}$) &$0.8 {L_{\rm d, 46}\over \eta_{-1} M_9 \beta_{\rm r}}$ 
&$8 {L_{\rm d, 39}\over \eta_{-1} M_1 \beta_{\rm r}}$ 
&$10^{12} {M_{\rm T, -1}\over  M_1 \beta_{\rm r} t_2 }$ &\\
 & & & & \\
$B_{\rm eq}$(@$R_{\rm s}$)  & $10^4  \left({L_{\rm d, 46} R_{\rm s}/h 
 \over (\eta_{-1} M_9 \beta_{\rm r} }\right)^{1/2}$  
                  & $4 \times 10^8 \left( {L_{\rm d, 39} R_{\rm s}/h 
 \over \eta_{-1} M_1 \beta_{\rm r} } \right)^{1/2}$
                  & $2 \times 10^{14} \left( {M_{\rm T,-1} R_{\rm s}/h 
 \over t_2 M_1^2 \beta_{\rm r} }\right)^{1/2}$  & G \\
 & & & & \\
\hline
\end{tabular}
\end{center} 
\end{table}

%

\section{Internal shocks}

In 1978, Rees proposed that the jet of M87 could be powered
by collisions among different part of the jet itself,
moving at different speeds.
When colliding, they would produce shocks, giving rise to the
non--thermal radiation we see.
Although born in the AGN field, this idea of {\it internal shocks}
grew up more robustly in the GRB field, to become the
``paradigm" to explain their prompt emission
(see e.g. Rees \& M\'esz\'aros 1994).
Faster and later shells can then catch up slower earlier ones,
dissipating part of their bulk kinetic energy into radiation.
However, all shells are and remain relativistic: after the collision
the merged shells move with a bulk Lorentz factor which is
intermediate between the two initial ones.
It is then clear that this mechanism has a limited efficiency, because
only a small fraction of the bulk energy can be converted into radiation
(unless the contrast between the two initial Lorentz factor is huge,
see Beloborodov 2000, but also Ghisellini 2002).

In blazars, on the other hand, we {\it require} a small efficiency,
since most of the bulk energy has to go un--dissipated to the
outer radio--lobes.
This model is very promising, since it can explain some basic properties
of blazars:

\begin{itemize}

\item About 10 per cent of the power in bulk motion is dissipated
into radiation, while 90 per cent is available to power the
radio lobes.

\item Most of the dissipation occurs at a few hundreds of
Schwarzschild radii, far enough from the black hole and accretion disk
to avoid relevant $\gamma$--$\gamma$ $\to$ $e^\pm$ absorption of
$\gamma$--rays, but still within (in powerful objects)
the broad line region, which provides seed photons for the
Compton scattering process (leading to a dominant high energy
emission).

\item The shells that have already collided once can collide again,
at larger distances from the black hole with reduced efficiency,
since the $\Gamma$--contrast of the colliding shells decreases.
This can explain why the luminosity of jets decreases (but
it does not vanish) with distance.
Fig. 1 shows the efficiency (i.e. the fraction of 
the luminosity emitted to the jet carried power) as a function
of distance for a simulation we have done for Mkn 421 (Guetta et al. 2002).

\item Variability is a built--in feature of the model. 
See Zhang et al. (2002, see also this volume), 
for an example of blazar variability
that might be explained by internal shocks.

\item There is the possibility (not yet studied in detail) of
a link between the flares at optical and $\gamma$--ray
energies and the flares in the radio--mm band (see Spada et al. 2001
for an example of correlation between the mm and the optical--$\gamma$--ray
fluxes in 3C 279).

\end{itemize}
\begin{figure}
\epsfysize=13cm 
\epsfbox{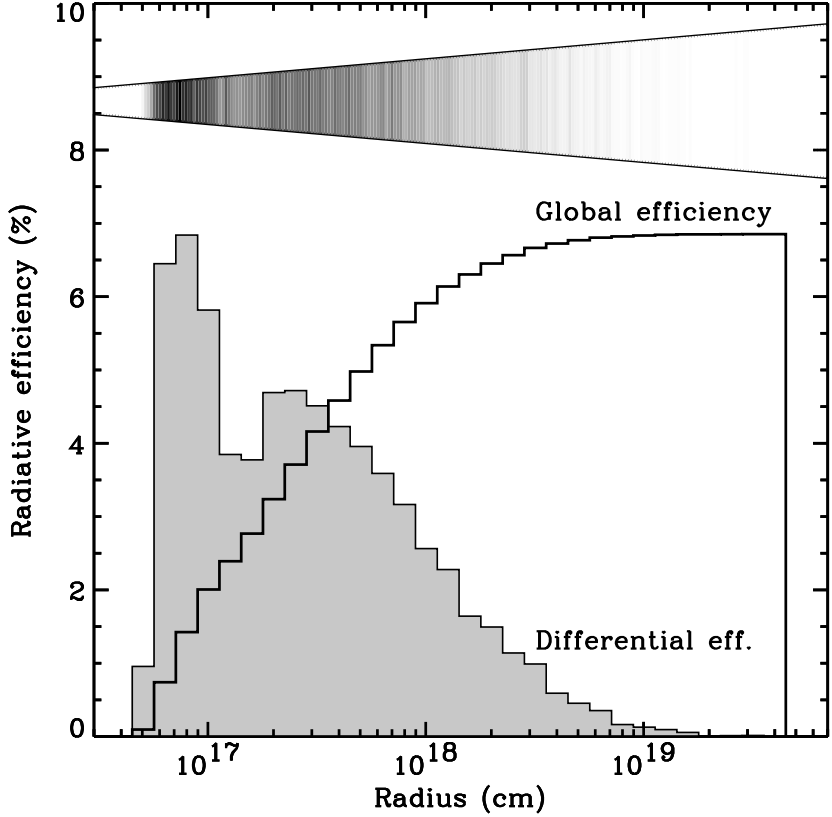}
\vskip -1 true cm
\caption[h]{The result of a simulation of the internal shock 
model for Mkn 421. 
We plot the 
average local radiative efficiency and the cumulative efficiency
as functions of distance from the black hole.
The first ``peak" in the efficiency
corresponds to the first collisions between shells, while the
``tail" at larger distances corresponds to second, third (and so on)
collisions. The minimum distance at which collision occur is dictated by
the initial separation of the shells $R_0$ and their average bulk Lorentz
factor $\Gamma$. 
For the first collisions we have $R_{\rm coll}\sim \Gamma^2 R_0$. 
The top of the plot shows a schematic representation of the jet,
with grey levels proportional to the local efficiency. From Guetta et al. 
(2002).
}
\end{figure}

A detail study via numerical simulations of the predictions of
this model in blazars has been carried out by Spada et al. (2001)
for powerful objects (like 3C 279).
We are now studying less powerful objects (like Mkn 421 and BL Lac itself,
Guetta et al. 2002).
As discussed by Kaiser, Sunyaev \& Spruit (2000),
internal shocks could also work for Galactic Superluminals,
although detailed simulations have yet to be done.

\subsection{Internal shocks and the blazar sequence}

One of the important characteristics of the internal shock model
for blazars is that the injection of energy in each collision
is finite in time.
The generation of relativistic particles lasts for the collision
time, which is of the order of the time needed for one shell to
overtake the other, i.e. $t_{\rm inj}^\prime \sim \Delta R^\prime/c$
in the comoving frame.
The energy distribution of the relativistic particles responsible 
for the emission, in this case, never reaches a steady state, and
the maximum emitted flux corresponds to the end of the injection 
(if the energy distribution of particles, $N(\gamma)$, is steeper than
$N(\gamma)\propto \gamma^{-2}$).
{\it This can explain why there is a sequence in the SED of blazars:}
in low power sources lacking emission lines the cooling time is long,
and only the most energetic particles can cool in $t_{\rm inj}$.
Suppose to inject a power law of relativistic particles, between
$\gamma_{\rm min}$ and $\gamma_{\rm max}$, with slope $s$
[i.e. $Q(\gamma)\propto \gamma^{-s}$].
After the injection phase (i.e. after $t_{\rm inj}$), only those particles
for which $t_{\rm cool}(\gamma)<t_{\rm inj}$ have cooled, producing
a break in the particle distribution at $\gamma_{\rm c}>\gamma_{\rm min}$.
Below $\gamma_{\rm c}$ the particle distribution
retains its original slope [$N(\gamma)\propto \gamma^{-s}$],
while above $\gamma_{\rm c}$ the distribution has a slope
steeper by one $N(\gamma)\propto \gamma^{-(s+1)}$.
If $2<s<3$ the peaks of the synchrotron and the inverse Compton spectra
are produced by electrons at $\gamma_{\rm c}$.
For powerful sources, on the other hand, electrons of almost all energies
can cool in a time $t_{\rm inj}$, due to the larger magnetic
field in their jet, but especially due to the contribution of the 
broad line photons which enhances the inverse Compton cooling rate.
In this case the particle distribution 
$N(\gamma)\propto \gamma^{-(s+1)}$ above $\gamma_{\rm min}$ and
$N(\gamma)\propto \gamma^{-2}$ below.

{\it Therefore sources will have different peak locations, according to 
the amount of cooling suffered during the 
time of the shell--shell collision.}
Calling $\gamma_{\rm peak}$ the random Lorentz factor of those
electrons emitting most of the radiation (i.e. emitting at the 
peaks of the SED), we find a correlation between $\gamma_{\rm peak}$
and the amount of radiative plus magnetic energy density $U$
(as measured in the comoving frame) which has {\it two} branches:
in powerful objects $\gamma_{\rm peak}=\gamma_{\rm min} \propto U^{-0.5}$,
while in weak lineless BL Lacs we have 
$\gamma_{\rm peak}=\gamma_{\rm c} \propto U^{-1}$
(Ghisellini, Celotti \& Costamante, 2002).

\section{Conclusions}

We have rather compelling indications that relativistic jets
in different systems
are produced via similar processes, and that also the way in which
they dissipate part of their kinetic energy into radiation may
be similar.
The matter outflowing in relativistic jets may be of the order 
of 1 per cent of the matter inflowing into the black hole.
If powered by spinning black holes, the value of the magnetic field
close to the horizon must be of the order of the gravitational 
energy density of the accreting matter, which can be much more of 
the gas or radiation pressure. 

There is now growing consensus that the massive black hole in
radio--loud AGNs exceeds $\sim 10^8 M_\odot$ (see e.g. Lacy et al. 2001; 
Ghisellini \& Celotti 2001).
If maximally spinning, its rotational energy is more than enough to
account for the energy contained in the radio--lobes
of radio--loud sources.
There may be a problem, however, in the mechanisms proposed so far
to extract this rotational energy.
They must be very efficient.
Indeed, jets can transport more power than that radiated 
by accretion disks: this is obviously true in GRBs, but it is also the case
for lineless BL Lac objects and Galactic Superluminal sources 
during major radio flares.
In fact, as already known at low accretion rates, 
the accretion disk luminosity could be lowered by the de--coupling 
of proton and electrons
(Rees et al. 1982, Narayan, Garcia \& McClintock 1997),
while, at high accretion rates, the optical depth
of the accretion disk is large enough to trap photons (Begelman 1979).
Jets seem not to be bound by these limitations.


\acknowledgements
We are grateful to ASI and the Italian MIUR (AC) for support.


\end{document}